\documentclass[sigconf]{acmart}
\usepackage{graphicx}
\usepackage{amsmath}
\usepackage{booktabs}
\usepackage{float} 
\usepackage[table]{xcolor} 

\begin{document}

\title{Rethinking Multi-objective Ranking Ensemble in Recommender System: From Score Fusion to Rank Consistency}

\author{Boyang Xia}
\email{xiaboyang.tech@gmail.com}
\affiliation{%
  \institution{Kuaishou Technology}
\city{Beijing}
  \country{China}
}
\author{Zhou Yu}
\email{yuzhou03@kuaishou.com}
\affiliation{
\institution{Kuaishou Technology}
\city{Beijing}
  \country{China}
}

\author{Zhiliang Zhu}
\email{zhuzhiliang@zuaa.zju.edu.cn}
\affiliation{%
  \institution{Unaffiliated}
  \city{Beijing}
  \country{China}
  }

\author{Hanxiao Sun}
\email{sunhanxiao03@kuaishou.com}
\affiliation{%
  \institution{Kuaishou Technology}
\city{Beijing}
  \country{China}
}

\author{Biyun Han}
\email{hanbiyun@kuaishou.com}
\affiliation{%
  \institution{Kuaishou Technology}
\city{Beijing}
  \country{China}
}

\author{Jun Wang}
\email{wangjun03@kuaishou.com}
\affiliation{%
 \institution{Kuaishou Technology}
\city{Beijing}
  \country{China}
 }

 \author{Runnan Liu}
 \email{liurunnan@kuaishou.com}
\affiliation{%
 \institution{Kuaishou Technology}
\city{Beijing}
  \country{China}
 }

\author{Wenwu Ou}
\email{ouwenwu@gmail.com}
\affiliation{%
  \institution{Kuaishou Technology}
  \city{Beijing}
  \country{China}
  }

\renewcommand{\shortauthors}{Xia et al.}

\begin{abstract}
The industrial recommender systems always pursue more than one business goals, \textit{e.g., } short term purchases and long term user activeness in e-commerce recommender systems. The inherent intensions between objectives pose significant challenges for ranking stage. A popular solution is to build a multi-objective ensemble (ME) model to integrate multi-objective predictions into a unified score. 
Although there have been some exploratory efforts, few work has yet been able to systematically delineate the core requirements of ME problem.
We rethink ME problem from two perspectives. From the perspective of each individual objective, to achieve its maximum value the scores should be as consistent as possible with the ranks of its labels. From the perspective of the entire set of objectives, an overall optimum can be achieved only when the scores align with the commonality shared by the majority of objectives. However, none of existing methods can meet these two requirements. First, popular methods are optimized by classification or regression tasks, which is misaligned with ranking task. Second, although these methods build complex user-specific structure for multi-objective score fusion, none of them consider modeling the underlying commonality across objectives. 

To fill this gap, we propose a novel multi-objective ensemble framework \textbf{HarmonRank} to fulfill both requirements. For rank consistency, we formulate rank consistency (AUC) metric as a rank-sum problem and make the model optimized towards rank consistency in an end-to-end differentiable manner. For commonality modeling, we change the original relation-agnostic ensemble paradigm to a relation-aware one in two steps. First, we capture the commonality among objectives with via self-attention mechanism. Second, we fuse relation-aware objective representations into the ensemble score in a user-specific manner.

Extensive offline experiments results on two industrial datasets and online experiments demonstrate that our approach significantly outperforms existing state-of-the-art methods. Besides, our method exhibits superior robustness to label skew situations which is common in industrial scenarios. The proposed method has been fully deployed in Kuaishou's live-streaming e-commerce recommendation platform with 400 million DAUs, contributing 2.6\% purchase gain.
\end{abstract}

\ccsdesc[500]{Information systems~Recommender systems.}

\keywords{Multi-objective ranking ensemble, Personalized recommendation, Learning to rank}


\maketitle
\section{Introduction}
\begin{figure}
  \centering
\includegraphics[width=0.5\textwidth]{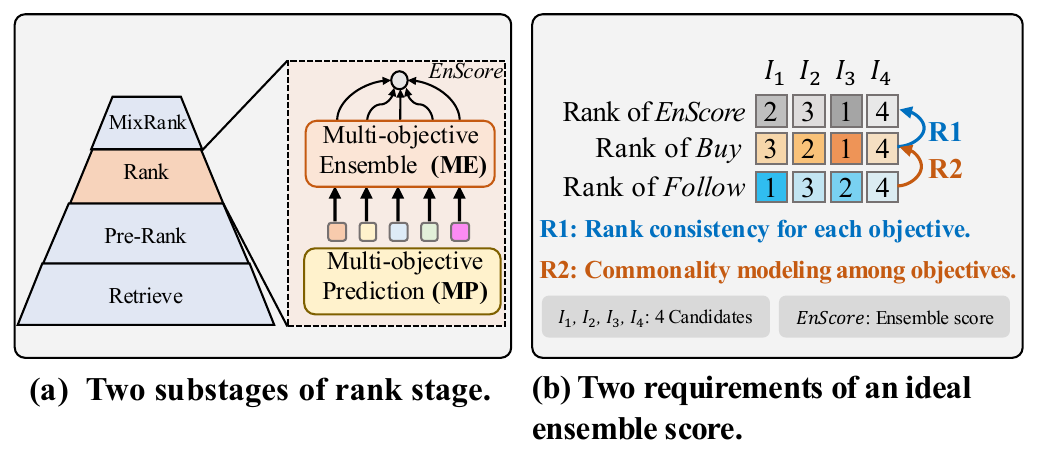}
  \vspace{-1em}
  \caption{Conceptual view of ranking stage and requirements of an ideal ensemble score.
  (a) Two sub-stages of ranking stage in industrial recommender systems, \textit{i.e.,} multi-objective prediction and multi-objective ensemble stages. (b) Two requirements of an ideal ensemble score. For each individual objective, the score should align with ranks of labels. For all objectives, the score need align with the majority of them.} 
  \label{overview1}
  \vspace{-1em}
\end{figure}


\textit{Background.} The industrial recommender systems always pursue more than one business goals. For instance, e-commerce recommender systems need to balance purchases (short-term gains) and user interactions (long-term ecosystem health). The inherent conflicts among these business goals pose significant challenges for multi-objective ranking stage in these recommender systems, \textit{e.g.,} straightforward pursuing purchase may reduce following behaviors of users, similar to exploit-exploration seesaw
problem. To balance these competing objectives, industrial ranking systems usually follow a two-stage paradigm (see Fig. \ref{overview1} (a)), \emph{i.e.,} multi-objective prediction (MP) and multi-objective ensemble (ME) \cite{ali_ltr,pantheon,intel,label_aggregation}. The MP stage is usually a high-capacity multi-task learning (MTL) model \cite{mmoe} whose goal is to predict the precise probability scores for multiple objectives, \emph{e.g.,} click through rate (CTR) and click-through conversion rate (CTCVR). Whereas ME stage is a light-weight model whose goal is to generate a unified ensemble score in a personalized manner. 
The ensemble score is used to truncate top items to represent for users. 
Although extensive research has been conducted on MP (MTL) problem in recent years, the ME problem remains more underexplored due to its strong dependence on industrial applications.
 
\begin{figure}
  \centering
\includegraphics[width=0.5\textwidth]{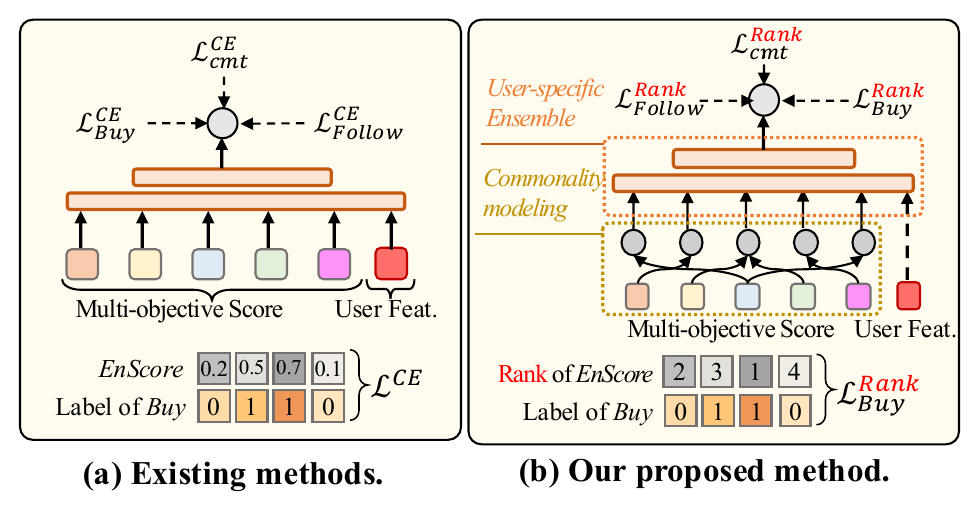}
  \vspace{-1em}
  \caption{Conceptual comparison between existing methods and our method. 
  (a) Existing methods. User feature conditioned multiple non-linear layers serve as the backbone and multiple binary CE losses serve as the training objectives.
 (b) Our methods based on two core design insights. (1) Ranking aligned optimization objective towards AUC. (2) Relation-aware architecture with two steps, \textit{i.e.} commonality modeling, and user-specific ensemble.} 
  \label{overview2}
  \vspace{-1em}
\end{figure}

 \textit{Motivation.} We ask, \textit{what requirements should an ideal ensemble score satisfy}? 
We answer this from two perspectives. On one hand, for each individual objective, to achieve its maximum value (\textit{e.g., } purchases), the scores must be as consistent as possible with the ranks of its labels (\textit{e.g., } labels of \textit{buy} behaviors). On the other hand, for the entire set of multiple objectives, an overall optimum can be achieved only when the scores align with the commonality shared by the majority of objectives. These imply two specific requirements for the design of ME models (see Fig. \ref{overview1} (b) for conceptual view).
\textbf{R1: Rank consistency.} For optimization objective, they should be capable of optimizing towards ranking consistency between ensemble scores and multi-objective labels directly. 
\textbf{R2: Commonality modeling.} For representation learning, they should be able to model the commonality among multiple objectives. 
However, we find that existing methods fail to meet these.
First, popular methods are optimized by classification or regression tasks \cite{intel,pantheon}. 
These tasks are misaligned with the ultimate ranking task (evaluated by AUC) \cite{tang2022smooth}. 
Second, although these methods build complex user-specific structure for multi-objective score fusion, none of them consider modeling the underlying commonality across objectives (see Fig. \ref{overview2} (a)). 




 
 
\textit{Our method.} To fill this gap, in this paper, we propose a novel personalized multi-objective ensemble framework \textbf{HarmonRank}, which enables both rank consistency and commonality modelling among multiple objectives (see Fig. \ref{overview2} (b)). 
For rank consistency, we firstly reformulate the non-differentiable AUC as a rank sum problem and utilize an advanced differentiable ranking technique to optimize multi-objective AUC in an end-to-end manner. For commonality modeling, we propose to change existing relation-agnostic ensemble architecture to a relation aware one by two steps. First, we align the common parts between multiple objectives and obtain relation-aware representations via self-attention mechanism. Then, we fuse the relation-aware representations of all objectives into the ensemble score in a user-specific manner. 

Overall, our contributions can be summarized as follows:
\begin{itemize}
    \item We propose a novel personalized multi-objective ensemble framework HarmonRank to enable consistency between optimization objective and the ranking task.
    \item We propose a relation-aware composite ensemble paradigm to capture the commonality between multiple objectives.
    \item We conduct extensive offline experiments on two industrial recommendation datasets. Experimental results demonstrate that our HarmonRank achieves superior performance over all existing state-of-the-art methods. Additionally, HarmonRank contributes 2.6\% purchase gain for Kuaishou live-streaming e-commerce platform. 
    \item We pioneeringly introduce a robustness analysis method for multi-objective ensemble model against label skew. We verify our method outperforms baselines in a large margin on robustness, which makes our method more universal in various industrial scenarios. 
\end{itemize}
\section{Related Work}
\subsection{Multi-objective optimization in recommender systems}
Multi-objective optimization is recognized as a key challenge in both research and industry community. Early recommender systems primarily optimized for only few objectives, \emph{e.g.} click-through rate (CTR) prediction for e-commerce recommender system and watch time for video recommender systems. However, with the increasing complexity user interface functionalities, morden recommender system need to balance more competing objectives. 
Some studies attend on the sample selection bias caused by chain relationships between objectives, \emph{e.g.,} CTR and conversion rate (CVR) \cite{essm,choruscvr}. Some other researchers are dedicated to improve multi-task learning mechanism between different objectives \cite{mmoe,ple}. For an example, Multi-gate Mixture-of-Experts (MMOE) \cite{mmoe} architecture addresses this by leveraging expert networks to capture shared representations across objectives, thereby improving accuracy of individual objectives.
Despite these advances, how to ensemble multiple prediction scores of different objectives remains a non-trivial challenge. Current industrial systems typically address this through a multi-objective ensemble model \cite{intel, unsupervised, ali_ltr}.
\subsection{Multi-objective Ensemble in recommender systems}
Early industrial approaches predominantly employed ranking formulas to combine scores from multiple objectives, typically optimizing parameters through grid search. While black-box optimization methods, \emph{e.g.,} Bayesian optimization \cite{bayesian} and cross entropy methods \cite{cross_entropy}, can reduce the iterations for searching optimum, they still suffer from two critical limitations: (1) inability to present personalized ranking mechanism to adapt user preferences, and (2) incapacity to exploit real-time user feedback. The factors matter for system since the optimal ensemble schemes change with different users and different time. To address these challenges, recent methods have proposed model-based ensemble solutions, to utilize more informative features and deep neural networks trained in hourly updated streaming feedback.

The key challenge for model-based multi-objective ensemble lies in the absence of a gold-standard supervision across multiple objectives. To solve this problem, some methods apply multiple unilateral binary classification losses to optimize towards all objectives simultaneously \cite{pantheon,peltr}, which can be referred as \textit{loss aggregation} methods. The other methods aim to build a unified supervision by aggregating multiple binary labels into a single regression label \cite{intel,label_aggregation}, which can be categorized as \textit{label aggregation} methods. Despite these efforts, the optimization directions of these methods' supervisions are misaligned with the ultimate ranking task (evaluated by AUC), which leads to suboptimal performance. Reinforce learning based methods \cite{ali_ltr} can directly optimize towards ranking task metrics (AUC sum), whereas their policy gradients suffer from large variance and instability. Few of existing methods consider to optimize final sum of AUC in an end-to-end differentiable manner. It is NP-hard to direct optimize AUC because of the non-convexity and discontinuousness  of \cite{auc_pl}.
\subsection{AUC optimization}
The AUC is a ranking quality metric defined based on pairwise comparisons \cite{auc,hanley1982meaning} under binary label settings. Thus, traditional differentiable AUC optimization methods concentrate on developing effective pairwise surrogate functions to approximate non-differentiable comparison function, \emph{i.e.,} $s_i - s_j \geq 0$. For examples, pairwise square methods \cite{psq} employ square functions $(1-(s_i-s_j))^2$ surrogate functions, while pairwise logistic methods \cite{auc_pl} propose to use logistic function $\text{log}(1+\text{exp}(-(s_i - s_j)))$ to approximate the comparison function. However, the inherent $O(n^2)$ time complexity of pairwise approaches renders them impractical for large-scale datasets. Consequently, numerous optimization attempts have shifted towards instance-wise AUC optimization. Leveraging the special form of pairwise square, pairwise AUC maximization can be decomposed into a min-max instance-wise formulation \cite{ying2016stochastic}, with $O(N)$ time complexity. Nevertheless, this instance-wise optimization introduces much sensitivity to noisy samples. To address this, AUCM \cite{aucm} proposed a margin-based min-max optimization strategy to effectively mitigates this issue.

While recent advancements improve the computational complexity to $O(N)$, they inadvertently introduce greater inconsistency with factual AUC computation due to computation simplification for surrogate functions. In this paper, to balance computational efficiency and AUC computation consistency, we propose a novel AUC  optimization method under a rank based formulation other than pairwise formulation. We formulate AUC computation as a rank-sum problem in the ordered score list and utlize differenable ranking technique to solve this rank-sum problem. Our method presents $O(n \text{log} n)$ time complexity and superior AUC optimization accuracy for alignment with AUC theoretical computation.
\section{Problem Formulation}

\begin{figure*}[t]
  \centering
\includegraphics[width=0.8\textwidth]{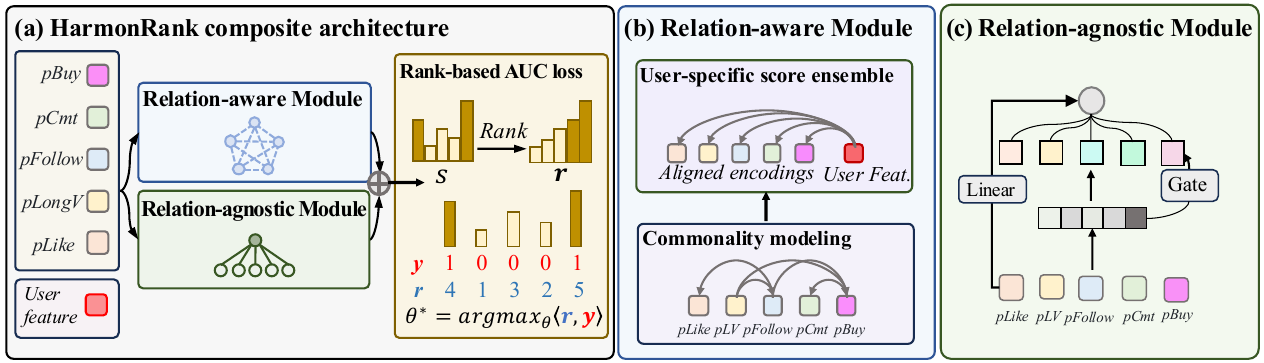}
  \caption{Systematic overview of our HarmonRank framewotk. (a) The architecture of HarmonRank. (b) Relation-aware module with inter-objective alignment and personalized guidance. (c) Relation-agnostic module with gate mechanism and linear fusion pathway. `pBuy', `pCmt', `pFollow', `pLV' and `pLike' here represent the predicted probability scores corresponding to `buy', `comment', `follow', `long view' and `like'.}
  \label{main}
  \vspace{-1em}
\end{figure*}
\label{sec:problem_formulation}
\subsection{Input and output} Let $\mathcal{I}_{cand}(\mathbf{u}, \mathbf{c})=\left\{I_i\right\}_i^N$ be a requested candidate set for the user $\mathbf{u}$ and environment context $\mathbf{c}$ (\emph{e.g.,} time and app version). For each item $I_i$, multi-objective prediction (MP) model estimates $M$ kind of interaction scores $\mathbf{e}_i\in\mathbb{R}^{[M,1]}$, \emph{e.g.,} purchase, long view, like, follow, \emph{etc}. Multi-objective ensemble (ME) model is responsible for integrating these scores into an ensemble score in a personalized way:
\begin{equation}
    s_i = \mathcal{F}(\mathbf{e}_i, \mathbf{u} , \mathbf{c} ;\theta)
 \end{equation}
 Then top items are selected and displayed to users based on top-k truncation of the ensemble scores $\mathcal{S}=\{s_i\}_i^N$, where $\mathcal{F}(\cdot;\theta)$ is a neural-network ensemble model parameterized by $\theta$.
\begin{equation}
    \mathcal{I}_{realshow} = \text{TopK}(\mathcal{I}_{cand}, \mathcal{S})
 \end{equation}
\subsection{Train and evaluation}
Similar to MP stage, we collect multiple behavior feedback labels in a fixed waiting window for each exposed sample in logs, and we use these labels to supervise the ensemble model training. Our goal is to achieve improved trade-offs across competing objectives. Due to the high cost of online AB metric evaluation, we typically evaluate the ME model's comprehensive ranking abilities on all objectives by the sum of AUC on a offline dataset $\mathcal{D} = \{(\mathbf{x}_i, \mathbf{y}_i)\}_{i=1}^N$. Here  $\mathbf{y}_i\in \{0,1\}^{[M,1]}$ is the ground truth interaction labels for the $i$-th sample:
\begin{equation}
\mathcal{G}_m(\mathcal{D}) = \frac{\sum_{s_i\in \mathcal{D}_m^+,s_j\in \mathcal{D}_m^-}\mathbb{I}(s_i \geq s_j)}{|\mathcal{D}_m^+|\cdot|\mathcal{D}_m^-|}
\end{equation} \begin{equation}
\mathcal{G}(\mathcal{D}) =\sum_{m=1}^{M} \mathcal{G}_m(\mathcal{D})
\end{equation}
where $\mathcal{D}_m^+$ and $\mathcal{D}_m^-$ represent the positive and negative sample sets defined on the $m$-th objective, \emph{e.g.,} the purchased items and unpurchased items. 
Existing methods primarily rely on classification or regression losses to supervise ME models, yet their optimization directions are substantial misaligned with the factual ranking ability (AUC). Therefore we propose an end-to-end ranking driven ME paradigm that explicitly optimizes the multi-objective AUC metric:
\begin{equation}
    \mathbf{\mathbf{\theta}}^* = \text{argmax}_{\mathbf{\mathbf{\theta}}}\mathcal{G}(\mathbf{\mathbf{\theta}})
    \label{eq:meta_optimization}
\end{equation}
\section{HarmonRank}
In this section we introduce how HarmonRank fulfill both two aforementioned requirements, rank consistency and commonality modeling. In Sec. \ref{sub1} we elaborate on rank consistency optimization. In Sec. \ref{sub2}, we elaborate on design of composite modules for commonality modeling.
\subsection{Rank consistency optimization}\label{sub1}
\subsubsection{Rank-based AUC formulation} 
Since $\mathcal{G}$ is non-differentiable w.r.t. $\mathbf{\theta}$, existing wisdom usually maximize AUC via surrogate function to approximate pairwise comparing function $\mathbb{I}(s_i \geq s_j)$ \emph{i.e.,} pairwise logistic \cite{auc_pl} and pairwise square \cite{psq} functions. However, this formulation brings large inconsistency between physical AUC computation. Meanwhile, pairwise computation introduce quadratic computation cost during training, which hinders further scaling on datasets.
To avoid these problems, we resort to alternative formulation for AUC computation.

In this section, we temporarily omit the subscripts $m$ of objectives for simplicity.
When sorting all positive and negative samples in an increasing order, AUC can be simplified to compute the rank sum of positive samples for the good property of ordered list \cite{rank_sum_test_v2}:
\begin{equation}
\mathcal{G}(\mathbf{w})=\frac{\langle\mathbf{r}, \mathbf{y}\rangle - |\mathcal{D}^+|\cdot(|\mathcal{D}^+|+1)/2}{|\mathcal{D}^+|\cdot|\mathcal{D}^-|}    
\end{equation}
where the vector $\mathbf{r}=\text{Rank}(\mathbf{s})$ represents the ranks of all $N$ samples and $\mathbf{y}$. It is noteworthy that, the conventional pairwise formulation suffers from cost inefficiency because of additional computations for positive-negative pairs with reverse orders ($s_{+} < s_{-}$). 
Under the ordered setting, AUC computation can be simplified to $O(nlogn)$ time complexity. 
To optimize $\mathcal{G}(\mathbf{w})$, we only need to optimize $\langle\mathbf{r}, \mathbf{y}\rangle$ for other terms are constant numbers. So the objective can be simplified as:
\begin{equation}
    \mathbf{\mathbf{\theta}}^* = \text{argmax}_{\mathbf{\mathbf{\theta}}}\langle\mathbf{r}, \mathbf{y}\rangle
    \label{eq:meta_optimization}
\end{equation}
However, for rank operation $\text{Rank}(\cdot)$ is non-convex, discontinuous gradients is infeasible for $\mathbf{s}$. To alleviate this, we employ a fast differentiable ranking algorithm \cite{fast_soft_sort} to enable gradient backpropagation from supervisions to the ensemble model.

\subsubsection{Differentiable Ranking}
Following \cite{fast_soft_sort}, we cast ranking operation (in increasing order here) as a discrete optimization problem over all feasible permutations $\Sigma$.
\begin{equation}
    \mathbf{r}^* =\text{argmax}_{\sigma\in \Sigma}\langle\mathbf{s}, \mathbf{r_\sigma}\rangle
\end{equation}
where $|\Sigma|=A^N_N$ and we desire the earlier-ranked samples should have smaller values. 
To transform this discrete optimization form to a continuous optimization problem, this method introduces a convex hull composed of all feasible permutations in $\Sigma$, which forms a  permutahedron $\mathcal{P}(\mathbf{r})$:
\begin{equation}
    \mathcal{P}(\mathbf{r}):=\text{conv}({r_\sigma}|\sigma\in\Sigma)
\end{equation}
To make it a convex objective, this method employs a quadratic regularization into the original optimization problem:
\begin{equation}
    P(\mathbf{s},\mathbf{r}) =\text{argmax}_{\mathbf{t} \in \mathcal{P}(\mathbf{r})}\langle\mathbf{s}, \mathbf{t}\rangle - \frac{1}{2}\parallel\mathbf{t}\parallel^2
    =\text{argmin}_{\mathbf{t}\in\mathcal{P}(\mathbf{r})}\frac{1}{2}\parallel \mathbf{t} - \mathbf{s} \parallel^2
\end{equation}
In this way, the original problem has been transformed to the projections to the permutahedron, which is a strong convex function. This approach enables forward propagation with $O(n \text{log} n)$ time complexity for sorting operation. In the same time, it cost only $O(n)$ time complexity during backward propagation for the gradients computation can be finished in the sorted sequence \cite{fast_soft_sort}.



\subsection{Commonality modeling} \label{sub2}
We find prior wisdom consistently overlook the commonality shared among multiple objectives. As shown in Fig. \ref{rela2}, there exist shared parts of varying degrees among objectives. To model the commonalities, we design a relation-aware composite architecture to integrate multiple scores in both relation-agnostic manner and relation-aware manner between objectives. As shown in Fig. \ref{main} (a), our framework is composed of two parallel components: relation-aware module and relation-agnostic module. The output scores of two modules are lately fused into a unified score to compute the AUC loss. 
\subsubsection{Relation-aware Module}
We align the shared parts between objectives with a relation-aware paradigm by two steps: first \textit{align} then \textit{ensemble}. For \textit{align} step, as shown in Fig. \ref{main}(b), we utilize self-attention mechanism \cite{self_attention} to capture pairwise relations between objectives. Through dynamic computation of relations between objectives, the aligned common parts between them are extracted. Specifically, the softmax normalized dot products between projected objective vectors, $\mathbf{Q}_r$ and $\mathbf{K}_r$, form the self attention matrix  $\mathbf{A}_{r} \in \mathbb{R}^{m\times m}$ between them. This matrix naturally represents the relation strengths between objectives. Then the original score embeddings are integrated into the representations $\mathbf{x}_r$ according to the attention matrix.
\begin{align}
\mathbf{Q}_r, \mathbf{K}_r, \mathbf{V}_r&=\mathbf{x}\mathbf{W}^{Q_r}, \mathbf{x}\mathbf{W}^{K_r}, \mathbf{x}\mathbf{W}^{V_r} \\
\mathbf{A}_{r} &=\text{Softmax}(\mathbf{Q}_s\mathbf{K}_s^{\top}/\sqrt{d_k})\\
\mathbf{x}_{r} &=\mathbf{A}_r\mathbf{V}_r
\end{align}
Although we capture inter-objective commonalities in this way, we still lack clear guidance on how to fuse multiple objectives into a unified score. Fortunately, user-specific personalized information including user profile and context feature, can serve as strong inductive bias to guide this fusion. We cast the personalized information as a \textit{query}, and search with it over multi-objective scores. We use the searched result as the fused score encodings to obtain the ensemble score. In this way, we implicitly let model answer a question that \textit{which objects can best represent the user’s holistic intentions in the current context?}. 
Specifically, we transform personalized information into the query vector $\mathbf{Q}_p$ by linear projection, similar to key and value vectors.
\begin{align}
\mathbf{Q}_p, \mathbf{K}_p, \mathbf{V}_p&=\mathbf{P}\mathbf{W}^{Q_p}, \mathbf{x}_r\mathbf{W}^{K_p}, \mathbf{x}_r\mathbf{W}^{V_p} \\
\mathbf{A}_{p} &= \text{Softmax}(\mathbf{Q}_p\mathbf{K}_p^{\top}/\sqrt{d_k})
\end{align}
Then attention weights $\mathbf{A}_{r} \in \mathbb{R}^{1\times m}$ , computed via dot product over all objective vectors $\mathbf{s}_r$, can serve as personalized guidance for importance of objectives for ensemble models. Then we fused score encodings by attention weights and transform the output to scalar ensemble score $\mathbf{s}_1\in\mathbb{R}$ by a linear projection layer. 
\begin{equation}
    \mathbf{s}_1 = \mathbf{w}_1^\top\left( \mathbf{A}_p \mathbf{V}_p \right) + b_1
\end{equation}
\subsubsection{Relation-agnostic Module}
Although we have effectively align the shared parts between different objectives, we still need to ensure that original information of each objective score can be fully preserved in final ensemble process. To this end, we introduce a relation-agnostic module (see Fig. \ref{main}(c)). In this module, we employ a gate mechanism to dynamically adjust the importance of different objectives based on their input scores. We obtain the gating coefficients $\mathbf{g}\in [0,1]^m$ by projecting score encodings through a linear layer with sigmoid activations. We control the score-wise retaining ratio of all score encodings through Hadamard product between coefficients  and score encodings. We then compute the output ensemble score by a final linear fusion layer over gated encodings.
\begin{align}
    \mathbf{g} &= \text{Sigmoid}(\mathbf{w}_g^\top\mathbf{x} + b_g) \\
    \mathbf{s}_2&=\mathbf{w}^\top_2(\mathbf{g} \odot \mathbf{x})+b_2
\end{align}
Simultaneously, to make ensemble model to learn a first-order linear fusion scheme, we maintain a parallel linear fusion pathway. In this way, we prevent model performance degradation when we add previous high-order compositions between objectives and make the whole ensemble model more robust.
\begin{align}
\mathbf{s}_3&=\mathbf{w}_3^\top\mathbf{x}+b_3
\end{align}
Finally, we obtain the output ensemble score by additive fusion of the outputs of both relation-aware module ($\mathbf{s}_1$) and relation agnostic module ($\mathbf{s}_2$ and $\mathbf{s}_3$).
\begin{align}
\mathbf{s}&=\mathbf{s}_1+\mathbf{s}_2+\mathbf{s}_3
\end{align}

\subsection{Pre-processing}
Before feed the score into relation-aware and relation-agnostic modules. We apply discretization embedding to represent the multi-objective scores to obtain non-linear representation. We use a simple equal distance discretization embedding technique to transform scalar scores to embeddings \cite{guo2021embedding}.
\begin{equation}
    \mathbf{e}^m=\mathbf{E}^m\cdot \text{floor}(\frac{\mathbf{s}^m}{B})
\end{equation}
where $B$ is the number of discrete buckets and $\mathbf{E}^m$ is the embedding matrix for the $m$th objective. The final inputs for the ranking model is the concatenation of both these discrete embeddings: $\mathbf{x}=[\mathbf{e}^1;...\mathbf{e}^m]$. In this way, we empower the model more non-linear fitting power comparing to simply using the scalar score representations.
 
\section{Experiments}
\begin{table*}[ht]
\resizebox{0.64\textwidth}{!}{
\begin{tabular}{l|ll|ll}
\toprule
       & \multicolumn{2}{c|}{TenRec-QKVideo}     & \multicolumn{2}{c}{Kuaishou-ELive}    \\ \hline
method & 3-objectives   & 5-objectives   & 3-objectives   & 5-objectives   \\ \hline
Multi-objective BCE \cite{pantheon}    & 2.2420 & 3.6465& 2.1555 & 3.6682\\
Label aggregation \cite{intel}    &  2.4040 & 3.8311& 2.0765 & 3.6883 \\
Reinforce learning \cite{ali_ltr}  & 2.2488 & 3.9074& 2.1930 & 3.7467 \\
Ours    & \textbf{2.4617} & \textbf{3.9186}& \textbf{2.2191} & \textbf{3.7522}\\ \bottomrule
\end{tabular}
}
\caption{Offline results on AUC sum on two industrial datasets under two settings. For TenRec-QKVideo, `3 objectives' are \textit{click}, \textit{like} and \textit{follow}, while `5 objectives' include \textit{click}, \textit{like}, \textit{follow}, \textit{share} and \textit{long view}. For Kuaishou-ELive, the `3 objectives' are \textit{buy}, \textit{follow} and \textit{long view}, while `5 objectives' include \textit{buy}, \textit{follow}, \textit{like}, \textit{comment} and \textit{long view}.}
\vspace{-2em}
\label{main_table}
\end{table*}
In this section, we conduct extensive online and offline experiments to verify the efficacy of our model following 4 research questions: 
\begin{itemize}
\item \textbf{RQ1:} How does our model perform on offline datasets?
\item \textbf{RQ2:} How does our module design and hyperparameter choice impact the performance?
\item \textbf{RQ3:} How does our method perform in practical training and inference runtime?
\item \textbf{RQ4:} Can our method bring improvements of A/B test metrics on online product environment?
\item \textbf{RQ5:} How our method perform in severe label skew situation? How our method impact trade-off between objectives? What does the model learn in multi-objective commonality modeling? 
\end{itemize}
\vspace{-1em}
\subsection{Implementation Details}
Before training ensemble model, we firstly prepare the input multi-objective scores via performing inference with a high-capacity ranking model over both datasets. For Kuaishou-Elive, we use the online in-service model trained incrementally by years of logs. For the public TenRec dataset, we train a standard MMOE \cite{mmoe} ranking model. Based on prepared multi-objective scores, all offline models are trained with 500 epochs with SGD optimizer. In terms of online training model with streaming data, it is usually trained for one epoch for the consideration of run-time cost and one-epoch overfiting risk in ranking stage. This scheme results in low sample efficiency. Since the model of ME stage is much more lightweight than MP stage and no sparse ID feature is used (cause of one-epoch overfitting), we train 20 epochs for each batch of streaming samples, to balance run-time cost and sample efficiency. We set learning rate to 1e-4 and batch size = 10240 for all methods. We train our method and compared methods with the same equal loss weights (all weights set to 1.0) for fair comparison on objective trade-offs. 
\vspace{-1em}
\subsection{Datasets}
To evaluate the performance of our method and compared baselines, we conduct comprehensive experiments on two real-world industrial datasets. 

\textit{Public dataset.} TecRec-QKVideo \cite{yuan2022tenrec} is a popular multi-objective recsys dataset collected from QQ-KAN video feeds platforms. It contains user interaction logs including five type of feedbacks, click, like, share, follow and long view. 
We sample 2M logs from all dataset, 1M for training, 0.5M for validation and test, respectively.
For the need of personalized ensemble, we utilize 2 user profile features, \textit{i.e.,} age, gender, \emph{etc.} 

\textit{Private dataset.} We build a new dataset called \textit{Kuaishou E-Live}, which contains both e-commerce feedback (purchase) and user engagement oriented feedbacks (long view, like, follow and comment). We built it by sampling millions of exposure logs at Kuaishou e-commerce platform, which is a pioneering live-streaming e-commerce platform with over 400 million DAUs. We sample 2M logs from one week, 1M for training, 0.5M for validation and test, respectively.  We utilize 4 context and user profile features, \textit{i.e.,} the hour of a day, app version, age and gender, to characterize user preference in ensemble learning. 

\vspace{-1em}
\subsection{Compared Methods.}
We compare our methods with most representative methods on multi-objective ensemble task. For fair comparison, we use the same network structure as HarmonRank on these baselines.
\begin{itemize}
    \item \textbf{Multi-objective binary cross entropy based (M-BCE)} \cite{pantheon}: This method enables optimization towards multiple objectives by integrate joint learning under multiple independent binary cross entropy losses for different objectives.
    \item \textbf{Reinforce learning (RL) based} \cite{ali_ltr} : RL-based method formulate multi-objective ensemble as a markov decision process, and treat input multi-objective scores as the `state' and the multi-object fusion score as the `action'.  The multi-object AUC over a batch of data serve as `reward'. 
    \item \textbf{Label aggregation based} \cite{label_aggregation,intel} : Label aggregation based method transforms multi-objective learning into single-objective learning by fusion multiple binary labels to a unified regression label. Then it utilize mean square error loss between the ensemble score and the aggregated label to optimize the ensemble model.
\end{itemize}


\subsection{Offline Results (RQ1)}
We evaluate the efficacy of our model by the multi-objective  AUC sum. We present the results on two settings, \emph{i.e.}, 3 objectives and all 5 objectives, to verify the conclusion consistency over different settings. For TenRec-QKVideo, `3 objectives' are \textit{click}, \textit{like} and \textit{follow}, while `5 objectives' include \textit{click}, \textit{like}, \textit{follow}, \textit{share} and \textit{long view}. For Kuaishou-ELive, the `3 objectives' are \textit{buy}, \textit{follow} and \textit{long view}, while `5 objectives' include \textit{buy}, \textit{follow}, \textit{like}, \textit{comment} and \textit{long view}.

The experimental results on two datasets are shown in Tab. \ref{main_table}. 
 On the TenRec-QKVideo dataset, our proposed method outperforms MBCE \cite{pantheon} by 27.2 pp (\textbf{3.919} \emph{v.s.} 3.647) for 5-objective AUC sum. On the Kuaishou-Elive dataset, our model outperforms MBCE by 8.4 pp (\textbf{3.7522} \emph{v.s.} 3.6682) for 5-objective AUC sum. The comparison results to the MBCE model demonstrates the efficacy of our proposed ranking-aligned ensemble model. 
 
The RL-based model is the best performing baseline on both datasets. On the TenRec-QKVideo dataset, our model outperforms RL-based model \cite{ali_ltr} by 1.2 pp (\textbf{3.919} \emph{v.s.} 3.907) for 5-objective AUC sum. On the Kuaishou-ELive dataset, our model outperforms RL-based model by 0.55 pp (\textbf{3.7522} \emph{v.s.} 3.7467). These results verify our continuous gradient-based method is superior to RL policy gradient based methods.
In a nutshell, our model consistently outperforms the best-performing baselines on both two datasets in a large margin in terms of different experimental settings. 

\begin{figure*}[t]
  \centering
\includegraphics[width=0.8\textwidth]{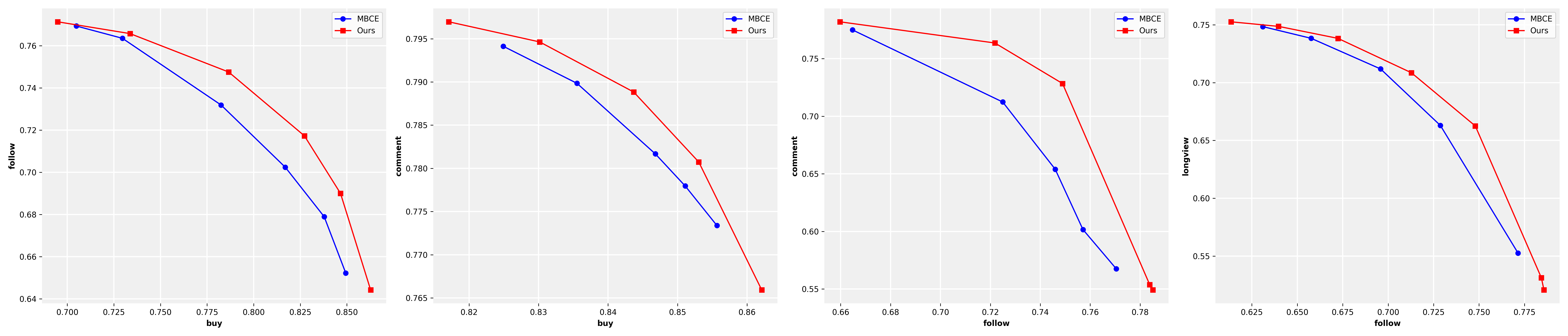}
  \caption{The tradeoff curves between competing objectives of the different models on the Kuaishou Elive dataset.}
  \label{pareto}
  \vspace{-1em}
\end{figure*}

\subsection{Ablation study (RQ2)} 
We conduct comprehensive ablation study on Kuaishou-ELive dataset to investigate the impact of different AUC optimization losses and different model structures.
\subsubsection{Different AUC optimization losses.}
Other than our proposed differentiable ranking based AUC maximization methed, we compare several popular pairwise AUC surrogate losses \emph{i.e.,} pairwise logistic \cite{auc_pl}, pairwise square \cite{psq} and the state-of-the-art instance-wise AUC maximization method AUCM \cite{aucm}. Pairwise methods maximize AUC objective with logistic or square function surrogate to approximate pairwise $\mathbb{I}(f(\mathbf{x}_+) \geq f(\mathbf{x}_-))$. For pairwise computation nature, these methods perform a $O(n^2)$ time complexity during training process. Different from them, AUCM transforms pairwise maximization objective to an equivalent instance-wise min-max problem, which reduce the training time complexity to $O(n)$. Notwithstanding these efforts, the inherent inconsistency with the AUC metric persists. We compare the practical training speeds of different losses in Section A of the Appendix.
\begin{table}[t]
\resizebox{0.46\textwidth}{!}{
\begin{tabular}{l|cc} 
\toprule
Different Losses&AUC Sum &Time-complexity\\
\hline
Pairwise Square &3.7415 &$O(n^2)$\\
Pairwise Logistic &3.7433 &$O(n^2)$\\
AUCM &3.7412 &$O(n)$\\
Ours (Differentiable Ranking) &\textbf{3.7522} &$O(n\text{log}n)$\\ 
\bottomrule
\end{tabular}
}
\caption{Results on Kuaishou-ELive for different AUC-driven losses.}
\label{ablation_aucloss}
\vspace{-3em}
\end{table}
As shown in Tab. \ref{ablation_aucloss}, our proposed differentiable ranking based method consistently outperforms other AUC losses. This is because our method eliminates the need of defining surrogate functions which makes it inevitably misaligned with AUC metric. This experiment fundamentally demonstrates the superiority of our method over existing methods.  
\subsubsection{Model components}
As shown in Tab. \ref{ab:component}, we conducted a comprehensive ablation study to evaluate the impact of different modules in our proposed method. We conduct three ablations in the relation aware module. First, when removing the self-attention-based alignment module (\texttt{Ours w/o Self-Attention}), we find the corresponding variant exhibits a 0.3pp drop in AUC sum (3.7494 \emph{v.s.} \textbf{3.7522}), indicating the indispensable role of relation modelling between different objective score encodings. Second, the variant without personalized feature (\texttt{Ours w/o Personalized Feat.}) exhibits a 0.6pp performance degradation (3.7464 \emph{v.s.}\textbf{3.7522}), demonstrating their significant contribution in guiding score importance weighting. Third, in terms of the manner of introducing personalized information, in addition to cross-attention mechanism, we experiment with an alternative baseline that simply concatenating personalized features with self-attention outputs before linear projection (\texttt{Ours w/o Cross-Attention}). Our method outperforms this baseline by 0.1pp in AUC sum (\textbf{3.7522 vs. }3.5712). We attribute this to linear projection's inability to facilitate dense interactions between personalized features and score encodings which can be fulfilled by cross-attention's dynamic weighting mechanism. 

In relation-agnostic module, we conduct ablations on two sub-components inside it. First, we try to remove the gate mechanism (\texttt{Ours w/o Gate Mechanism}), which leads to substantial performance drop (1.6pp), confirming its critical function for denoising and importance score selection. In addition, the variant without linear fusion pathway \texttt{Ours w/o Linear} results in a 0.3pp AUC sum decrease (3.7493 \emph{v.s.} \textbf{3.7522}), suggesting that the first-order linear fusion can effectively enhance model robustness.
\begin{table}[t]
\resizebox{0.3\textwidth}{!}{
\begin{tabular}{ll} 
\toprule
Method & AUC Sum\\\hline
 \multicolumn{2}{l} {\cellcolor{gray!20}\textit{In relation-aware module}}\\ \hline
Ours w/o Self-Attention& 3.7494\\
 Ours w/o Personalized Feat.&3.7464\\ 
 Ours w/o Cross-Attention &3.7512\\ \hline 
 \multicolumn{2}{l}{\cellcolor{gray!20}\textit{In relation-agnostic module}}\\\hline
 Ours w/o Gate Mechanism & 3.7362\\
Ours w/o LinearPath & 3.7493\\
Ours& \textbf{3.7522}\\ 
\bottomrule
\end{tabular}
}
\caption{Results on Kuaishou-Elive for different variants of our proposed structure with different components removed.}
\vspace{-2em}
\label{ab:component}
\end{table}
\subsubsection{Pre-processing.}
We also conduct experiments on different choices of number of buckets for score discretization. As shown in Tab. \ref{ab:dict_size}, we find the variant without discretization (\texttt{No Disc.}) results into a significant performance decrease (3.7522 $\rightarrow$ 3.7470), which demonstrates the necessity of discretization. We find different number of buckets present different performance, this may be because excessively fine-grained discretization approach the no-discretization situation, whereas overly coarse discretization make it lack discrimination between high and low scores. An appropriately number of buckets, \emph{i.e.,} 300 here, yields optimal performance.
\begin{table}[t]
\resizebox{0.3\textwidth}{!}{
\begin{tabular}{c|l} 
\toprule
\#DiscretizationBuckets & AUC Sum\\
\hline
 No disc.&3.7470\\
100& 3.7507\\
 300&\textbf{3.7522}\\
 600&3.7492\\
 \bottomrule 
\end{tabular}
}
\caption{Results on Kuaishou-Elive for different number of discretization buckets.}
\vspace{-2.5em}
\label{ab:dict_size}
\end{table}
\subsection{Practical runtime (RQ3)}
\textbf{Training.} As shown in Tab. \ref{tab:train_speed}, we compare the training speed among different AUC end-to-end losses. The training batch size is set to 10240 for all methods. The ranks of actual training speeds across different methods aligns with their theoretical computational complexity. Due to the pairwise nature of `PL' (Pairwise Logistic), it exhibits the slowest training speed, while AUCM and MBCE methods achieve the fastest training owing to their instance-wise loss computation with $O(n)$ complexity. Although our method's training speed lies between pairwise and instance-wise approaches, it delivers the best AUC performance across all methods.
\begin{table}[t]
    \centering
    \resizebox{0.4\textwidth}{!}{
    \begin{tabular}{c|c|c}
    \toprule
        Method & Practical training speed & Theoritical time complexity   \\ \midrule
        MBCE & 27.4K/s & $O(n)$   \\ \midrule
        PL & 24.8K/s & $O(n^2)$   \\ \midrule
        AUCM & 26.9K/s & $O(n)$   \\ \midrule
        HarmonRank & 26.0K/s & $O(nlogn)$ \\ \bottomrule
    \end{tabular}
    }
    \caption{Comparison of practical training speed.}
    \vspace{-2em}
    \label{tab:train_speed}
\end{table}
\textbf{Inference.}
We test our practical runtime during serving on an Intel 6230R CPU with  NVIDIAT4 GPU. 
The testing batchsize is 10240. As shown in Tab. \ref{tab:infer_speed}, our method costs negligible FLOPs and parameters, which is very appropriate for multi-objective ensemble stage.
\begin{table}[t]
    \centering
    \resizebox{0.4\textwidth}{!}{
    \begin{tabular}{c|c|c|c|c}
    \toprule
        FLOPs & Params & GPUMem & Latency & Throughput  \\ \midrule
        273M & 2.4K & 2.0GB & 4ms & 2.5M/s \\ \bottomrule
    \end{tabular}
    }
    \caption{Practical inference cost of our model.}
    \label{tab:infer_speed}
    \vspace{-2em}
\end{table}

\subsection{Online Results (RQ4)}
\begin{figure}[t]
  \centering
\includegraphics[width=0.4\textwidth]{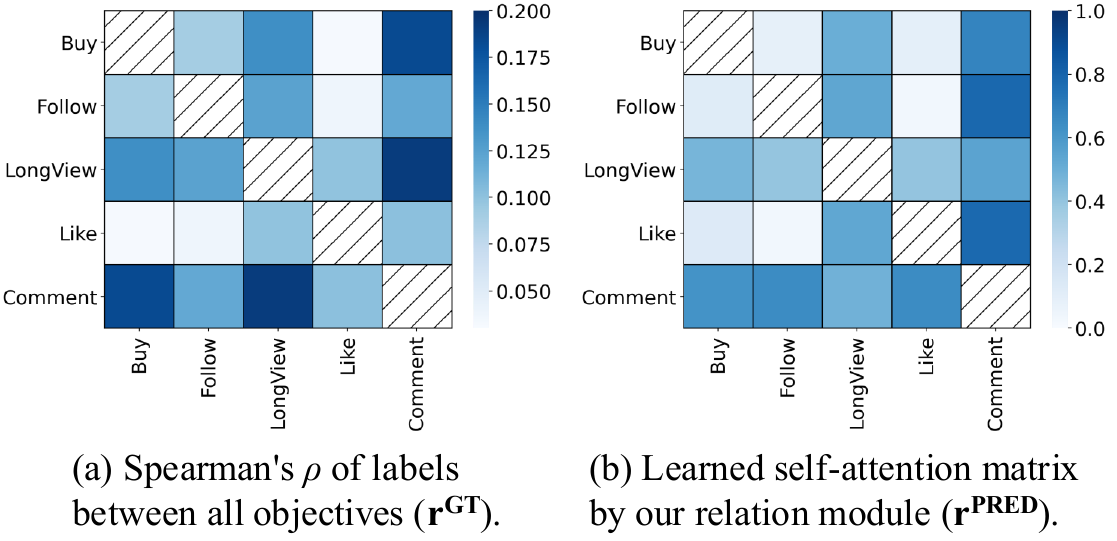}
  \caption{Consistency analysis between the learned attention matrix $\mathbf{r}^{PRED}$ and ground-truth data distribution $r^{GT}$ (label's Spearman rank correlation.}
  \label{rela2}
\end{figure}
We deploy our method in the production environment of Kuaishou e-commerce live-streaming platform to conduct online A/B testing for 5 days. 
It is noteworthy that the online base model is MBCE.
Compared with the base model. As shown in Tab. \ref{online} our model achieves significant improvements across all objectives, \emph{e.g.,} 2.635\% on core purchases metric and 0.451\% on follow metrics, which verify the effectiveness of our proposed HarmonRank for offline-online consistent improvements.
\begin{table}[t]
\centering
\resizebox{0.25\textwidth}{!}{
    \begin{tabular}{c|cl}
    \toprule
        Objectives  & RL&Ours\\ \midrule
        Purchase & +1.547\%  &\textbf{+2.635\%} \\ 
        Comment & -0.253\%  &\textbf{+3.034\%} \\ 
        Follow & +0.302\%  &\textbf{+0.451\%} \\ 
        Watch time & -0.238\%  &\textbf{+0.290\%} \\
        Like  & +0.817\%  &\textbf{+1.673\%} \\
    \bottomrule
    \end{tabular}
    }
    \caption{Online A/B Test Performance.}
    \vspace{-3em}
\label{online}
\end{table}

\subsection{Analysis (RQ5)}
\subsubsection{Robustness to label skew.}
To validate the robustness of our method against label skew, we conduct extensive experiments on the Kuaishou E-live dataset. Among our five objectives, the \textit{buy} objective exhibits the most severe label skew with a positive-to-negative ratio of $1:10^3$. To explore the performance of different methods under harder situations, we artificially strengthen the imbalance of \textit{buy} objective by down-sampling positive samples to ratios of $1:10^4$ and $1:10^5$ respectively on training set only, remaining the test set unchanged. 

As shown in the Tab. \ref{tab:label_skew}, when facing increasingly severe label skew, all methods exhibit expected AUC degradation. Nevertheless, our HarmonRank consistently maintains superior performance across all scenarios. Crucially, the performance drop of our method (-0.3\%) is significantly smaller than those of MBCE and PL (9.8\% and 4.3\%), demonstrating our methods has exceptional robustness to label distribution skew. 
     We attribute this phenomenon to different degrees of dependable on negative samples of these methods. 
As positive samples become sparser, classifying negative samples becomes increasingly trivial, resulting in diminishing gradients from the overwhelming majority of easy negative examples. 
Consequently, methods like MBCE and PL that heavily rely on negative samples suffer significant performance degradation. In contrast, HarmonRank inherently rely less on negative samples for its rank-sum formulation, whose loss only computes ranks of positive samples and negative samples are utilized only during the differentiable ranking phase. 
\begin{table}[t]
    \centering
    \resizebox{0.48\textwidth}{!}{
    \begin{tabular}{c|c|c|c}
    \toprule
        Methods & Original ($1:10^3$) & $10\times$ Skew ($1:10^4$) & $100\times$ Skew ($1:10^5$) \\ \midrule
        MBCE & 3.67 & 3.48 (-5.2\%$\downarrow$) & 3.31 (-9.8\%$\downarrow$) \\ \midrule
        PL & 3.74 & 3.60 (-3.7\%$\downarrow$) & 3.58 (-4.3\%$\downarrow$) \\ \midrule
        \textbf{HarmonRank} & \textbf{3.75} & \textbf{3.74 (-0.3\%$\downarrow$)} & \textbf{3.70 (-1.3\%$\downarrow$)} \\ \bottomrule
    \end{tabular}
    }
    \caption{Robustness against label skew.}
    \label{tab:label_skew}
    \vspace{-2em}
\end{table}

\subsubsection{Impact on Pareto frontier.}
To validate whether our method can achieve Pareto improvement (i.e., improving at least one objective without deteriorating others) in multi-objective optimization, we compare the Pareto frontiers between our proposed method and Multi-objective BCE. Given that the loss weights assigned to different objectives significantly influence their corresponding AUC performance, we systematically adjusted the allocation of loss weights to obtain a series of AUC values under various trade-off scenarios.
As seen in Fig. \ref{pareto}, our method consistently achieves superior trade-off curves over Multi-objective BCE (MBCE). Taking \textit{buy} and \textit{follow} objectives as examples, our method outperforms MBCE in terms of AUC for both objectives across all weight allocations. The similar phenomenon can be observed in trade-off curves between other objective pairs, including \textit{comment v.s. buy}, \textit{comment v.s. follow} and \textit{long view v.s. follow} in Fig. \ref{pareto}.

\subsubsection{Inter-objective relation analysis.}
To shed light on the factual commonalities among objectives (ground-truth), and which of them can be learned in inter-objective relations modules (predictions), we visualize inter-objective relations in two ways. For the ground-truth relations, we compute Spearman's rank correlation coefficients $\mathbf{\rho}$ between different objectives over the labels, which measures the monotonic correlations between the ranks of two objective labels. 
\begin{align}
r_{mn}^{GT} &= \text{Spearmanr}(\mathbf{y}_m, \mathbf{y}_n)
\end{align}
For predicted relations, we visualize the attention weights inferred by our method's relation module. 
\begin{align}
r_{mn}^{PRED} &= \mathbf{A}_r[m, n]    
\end{align}
As shown in Fig .\ref{rela2}, we can observe remarkable similarity between learned attentions $\mathbf{r}^{PRED}$ and empirical label correlations $\mathbf{r}^{GT}$, confirming our method's capability to accurately capture inter-objective commonalities.


\section{Conclusion}
The proposed HarmonRank framework addresses two limitations in multi-objective ensemble problem in recommendation systems. For the lack of alignment to ranking, we formulate AUC computation as rank-sum problem and use differentiable ranking to enable end-to-end AUC optimization. For the lack of alignment between different objectives, we propose a two-step paradigm, first \textit{align} then \textit{ensemble}. This paradigm effectively align the shared parts between objectives. Extensive online and offline experimental results demonstrate significant improvements of our proposed HarmonRank over existing methods.
\newpage
\bibliographystyle{ACM-Reference-Format}
\bibliography{sample-base.bib}
\clearpage 
\appendix 
\onecolumn 
\section*{Appendix} \label{appendix}
In this appendix, we provide more experimental results of our proposed HarmonRank. Accordingly, we organize the appendix as follows.
\begin{itemize}
    \item In Section \ref{appendix:runtime}, we present practical training speed our method and compared methods.
    \item In Section \ref{appendix:robustness}, we report the robustness of different methods under label skew situations. 
\end{itemize}

\section{Practical training speed}\label{appendix:runtime}
As shown in Tab. \ref{tab:train_speed}, we compare the training speed among different AUC end-to-end losses. The training batch size is set to 10240 for all methods. The ranks of actual training speeds across different methods aligns with their theoretical computational complexity. Due to the pairwise nature of `PL' (Pairwise Logistic), it exhibits the slowest training speed, while AUCM and MBCE methods achieve the fastest training owing to their instance-wise loss computation with $O(n)$ complexity. Although our method's training speed lies between pairwise and instance-wise approaches, it delivers the best AUC performance across all methods.
\begin{table}[H]
    \centering
    \resizebox{0.6\textwidth}{!}{
    \begin{tabular}{c|c|c|c}
    \toprule
        Method & Practical training speed & Theoritical time complexity & AUC Sum  \\ \midrule
        MBCE & 27.4K/s & $O(n)$ & 3.6682  \\ \midrule
        PL & 24.8K/s & $O(n^2)$ & 3.7433  \\ \midrule
        AUCM & 26.9K/s & $O(n)$ & 3.7412  \\ \midrule
        HarmonRank & 26.0K/s & $O(nlogn)$ & 3.7522\\ \bottomrule
    \end{tabular}
    }
    \caption{Comparison of practical training speed.}
    \label{tab:train_speed}
\end{table}
\section{Robustness}\label{appendix:robustness}
    To validate the robustness of our method against label skew, we conducted extensive experiments on the Kuaishou E-live dataset. 
    Among our five objectives, the \textit{buy} objective exhibits the most severe label skew with a positive-to-negative ratio of $1:10^3$. To explore the performance of different methods under harder situations, we artificially strengthen the imbalance of \textit{buy} objective by down-sampling positive samples to ratios of $1:10^4$ and $1:10^5$ respectively on training set only, remaining the test set unchanged. 

    As shown in the Tab. \ref{tab:label_skew}, when facing increasingly severe label skew, all methods exhibit expected AUC degradation. Nevertheless, our HarmonRank consistently maintains superior performance across all scenarios. Crucially, the performance drop of our method (-0.3\%) is significantly smaller than those of MBCE and PL (9.8\% and 4.3\%), demonstrating our methods has exceptional robustness to label distribution skew. 
    
    We attribute this phenomenon to different degrees of dependable on negative samples of these methods. 
    As positive samples become sparser, classifying negative samples becomes increasingly trivial, resulting in diminishing gradients from the overwhelming majority of easy negative examples. 
    Consequently, methods like MBCE and PL that heavily rely on negative samples suffer significant performance degradation.
    In contrast, HarmonRank inherently rely less on negative samples for its rank-sum formulation, whose loss only computes ranks of positive samples and negative samples are utilized only during the differentiable ranking phase. 
\begin{table}[!ht]
    \centering
    \begin{tabular}{c|c|c|c}
    \toprule
        Methods & Original ($1:10^3$) & $10\times$ Skew ($1:10^4$) & $100\times$ Skew ($1:10^5$) \\ \midrule
        MBCE & 3.67 & 3.48 (-5.2\%$\downarrow$) & 3.31 (-9.8\%$\downarrow$) \\ \midrule
        PL & 3.74 & 3.60 (-3.7\%$\downarrow$) & 3.58 (-4.3\%$\downarrow$) \\ \midrule
        \textbf{HarmonRank} & \textbf{3.75} & \textbf{3.74 (-0.3\%$\downarrow$)} & \textbf{3.70 (-1.3\%$\downarrow$)} \\ \bottomrule
    \end{tabular}
    \caption{Robustness against label skew.}
    \label{tab:label_skew}
\end{table}

\end{document}